%% file: main.tex
\PassOptionsToPackage{table}{xcolor}
\documentclass[sigconf,natbib=false,authorversion=true,nonacm=true]{acmart}

\ccsdesc[500]{Security and privacy~Systems security}
\ccsdesc[500]{Security and privacy~Key management}
\ccsdesc[300]{Computing methodologies~Modeling and simulation}
\ccsdesc[300]{Networks~Network simulations}

\keywords{Systems security; Satellite communication; Public key infrastructure; Network simulation; Delay tolerant networking}

\usepackage[utf8]{inputenc}
\usepackage{graphicx}
\usepackage{subcaption}
\usepackage{balance}
\usepackage[inline]{enumitem}
\usepackage{comment}
\usepackage{booktabs}
\usepackage{amsmath}
\usepackage{amsfonts}
\usepackage{url}
\usepackage{array}
\usepackage{makecell}
\usepackage{booktabs}
\usepackage{amsthm}
\usepackage[most]{tcolorbox}
\tcbuselibrary{theorems}
\definecolor{OxfordBlue}{HTML}{002147}

\usepackage[style=numeric-comp, sorting=none, backend=biber, maxbibnames=99]{biblatex}
\newtheoremstyle{colonit}{}{}{\itshape}{}{\bfseries}{:}{ }{}
\theoremstyle{colonit}

\newtcolorbox{hypothesis}[1][]{%
  colback=white, colframe=OxfordBlue,
  boxrule=0.5mm, arc=2mm,
  enhanced, breakable,
  left=8pt, right=8pt, top=6pt, bottom=6pt,
  title={Hypothesis\if\relax\detokenize{#1}\relax\else: #1\fi},
  fonttitle=\bfseries,
  before upper=\itshape
}

\usepackage{tikz}
\usetikzlibrary{calc,positioning,arrows}
\definecolor{cbsky}{HTML}{4891DC}
\definecolor{cbred}{HTML}{C10028}

\usepackage{pifont}
\newcommand{\cmark}{\ding{51}}%

\usepackage{siunitx}
\DeclareSIUnit[group-minimum-digits=3]\usd{USD}
\DeclareSIUnit[quantity-product = ]\percent{\char`\%}

\newcommand{\autoresizebox}[2]{%
    \centering
    \resizebox{%
        \ifdim#1\width>\linewidth
            \linewidth
        \else
            #1\width
        \fi
    }{!}{#2}%
}
\newcommand{\autobox}[1]{%
    \centering
    \resizebox{%
        \ifdim\width>\linewidth
            \linewidth
        \else
            \width
        \fi
    }{!}{#1}%
}

\newcommand{\sysname}[0]{\textsc{DSNS}}

\newcommand{\framework}[0]{\textsc{Key\-Space}}

\definecolor{tmpgreen}{HTML}{002147}
\definecolor{tmpred}{HTML}{C10028}
\colorlet{lightgreen}{tmpgreen!30}
\colorlet{lightred}{tmpred!20}
\newcommand{\best}[1]{\cellcolor{lightgreen} #1}
\newcommand{\worst}[1]{\cellcolor{lightred} #1}

\begin{document}

\title[\textsc{KeySpace}: Enhancing PKI for Interplanetary Networks]{\textsc{KeySpace}: Enhancing Public Key Infrastructure\\for Interplanetary Networks}

\author{Joshua Smailes}
\affiliation{%
  \institution{University of Oxford}
  \city{Oxford}
  \country{United Kingdom}
}
\email{joshua.smailes@cs.ox.ac.uk}

\author{Filip Futera}
\affiliation{%
  \institution{University of Oxford}
  \city{Oxford}
  \country{United Kingdom}
}
\email{filip.futera@cs.ox.ac.uk}

\author{Sebastian K{\"o}hler}
\affiliation{%
  \institution{University of Oxford}
  \city{Oxford}
  \country{United Kingdom}
}
\email{sebastian.kohler@cs.ox.ac.uk}

\author{Simon Birnbach}
\affiliation{%
  \institution{University of Oxford}
  \city{Oxford}
  \country{United Kingdom}
}
\email{simon.birnbach@cs.ox.ac.uk}

\author{Martin Strohmeier}
\affiliation{%
  \institution{armasuisse Science+Technology}
  \city{Z\"{u}rich}
  \country{Switzerland}
}
\email{martin.strohmeier@armasuisse.ch}

\author{Ivan Martinovic}
\affiliation{%
  \institution{University of Oxford}
  \city{Oxford}
  \country{United Kingdom}
}
\email{ivan.martinovic@cs.ox.ac.uk}

\renewcommand{\shortauthors}{Smailes et al.}

\input{abstract.tex}

\maketitle

\input{motivation.tex}

\input{background.tex}

\input{threat-model.tex}

\input{framework-design.tex}
\input{experiment-design.tex}

\input{results.tex}

\input{discussion.tex}

\input{conclusion.tex}

\section*{Acknowledgments}
The authors would like to thank Antonis Atlasis (ESA) for his insightful feedback which helped shape the paper, and armasuisse Science + Technology for providing funding and support for the project.
Joshua was supported by the Engineering and Physical Sciences Research Council (EPSRC).
Sebastian was supported by the Royal Academy of Engineering and the Office of the Chief Science Adviser for National Security under the UK Intelligence Community Postdoctoral Research Fellowships programme.
Simon was supported by the Government Office for Science and the Royal Academy of Engineering under the UK Intelligence Community Postdoctoral Research Fellowships scheme.

\printbibliography

\clearpage
\appendix
\input{appendices/experiment-configuration.tex}
\input{appendices/protocols.tex}
\input{appendices/data-processing.tex}
\input{appendices/latency-results.tex}
\input{appendices/revocation-coverage.tex}
\input{appendices/parameter-selection.tex}

\end{document}

%% file: abstract.tex
\begin{abstract}
As the use of satellites continues to grow, new networking paradigms are emerging to support the scale and long distance communication inherent to these networks.
In particular, interplanetary communication relays connect distant network segments together, but result in a sparsely connected network with long-distance links that are frequently interrupted.
In this new context, traditional Public Key Infrastructure (PKI) becomes difficult to implement, due to the impossibility of low-latency queries to a central authority.
This paper addresses the challenge of implementing PKI in these complex networks, identifying the essential goals and requirements.

Using these requirements, we develop the \framework{} framework, comprising a set of standardized experiments and metrics for comparing PKI systems across various network topologies, evaluating their performance and security.
This enables the testing of different protocols and configurations in a standard, repeatable manner, so that improvements can be more fairly tested and clearly demonstrated.
We use \framework{} to test two standard PKI protocols in use in terrestrial networks (OCSP and CRLs), demonstrating for the first time that both can be effectively utilized even in interplanetary networks with high latency and frequent interruptions, provided authority is properly distributed throughout the network.
Finally, we propose and evaluate a number of novel techniques extending standard OCSP to improve the overhead of connection establishment, reduce link congestion, and limit the reach of an attacker with a compromised key.
Using \framework{} we validate these claims, demonstrating their improved performance over the state of the art.
\end{abstract}

%% file: motivation.tex
\section{Motivation}\label{sec:motivation}

\begin{figure}
    \centering\includegraphics[width=\columnwidth]{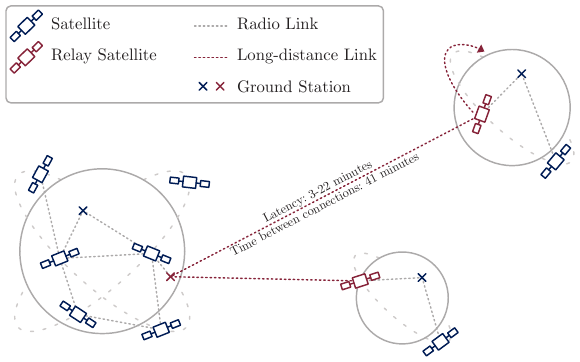}
    \caption{Example of communication across an interplanetary satellite network. There is a federated constellation of satellites around Earth (left), and relays connect the network to the Moon and Mars (right). These links are high-latency and, due to satellite and planet movement, are not always available.}
    \label{fig:overview-diagram}
\end{figure}

As space systems become an increasingly critical component of global infrastructure, satellite networks are both growing rapidly and becoming increasingly connected to each other.
This has been driven by recent advancements including improved launch technology, shared network infrastructure, CubeSat technology, and multi-tenant missions.
Additionally, interest in communication beyond Earth is resulting in greater numbers of satellites around the Moon, Mars, and in deeper space~\cite{israelLunaNet2020,nasascienceMars}.
As these networks grow, space agencies and other organizations are supporting them through the introduction and expansion of relay networks, building on existing networks like NASA's TDRSS and ESA's EDRS~\cite{monaghanTracking2015,esaEDRS2022}, with the introduction of lunar networks like NASA's LunaNet and ESA's Moonlight~\cite{israelLunaNet2020,nasaLunaNet2022,europeanspaceagencyMoonlight2024}.
In addition to these, the Mars Relay Network provides connectivity between Mars rovers and Earth~\cite{nasaMars}.
As these networks expand, they will result in an architecture composed of distinct connected segments, each partitioned by high-latency relay links.
On top of this, links are not always available due to orbital movement and occlusion by planets or the sun -- Figure~\ref{fig:overview-diagram} demonstrates this.

In this emerging landscape, one core challenge is to ensure the confidentiality and integrity of the communication.
Similar to terrestrial networks, one approach is to use Public Key Infrastructure (PKI).
However, space communications suffer from limited bandwidth and high transmission delays, which makes traditional PKI difficult to implement.
It is not straightforward to query a single centralized server for up-to-date information, and broadcast updates will propagate slowly throughout the network -- this can result in slow connection establishment or out-of-date certificate information.
Sometimes it is even impossible to form a connection between planets in the first place: every \num{26}~months the line of sight between Earth and Mars is interrupted by the Sun for two weeks~\cite{morabitoCommunications2001,nasaMars2015}.

This issue has been acknowledged by the Consultative Committee for Space Data Systems (CCSDS), who recently published the first experimental specifications of the Intergovernmental Certification Authority (IGCA), which states that ``certificate updates and certificate revocations are two of the most critical operations to maintain security.
[...] Because of limited contact periods, low bandwidth links, limited memory, and limited processor capabilities in comparison to most terrestrial systems, update rates and processes may vary from what is the norm for terrestrial certificates''~\cite{igcs_ccsds_357}.
It is clear that common approaches do not work in such systems and that it needs to be better understood how these challenges can be addressed.
A number of existing works have investigated key management in these and similar networks, but the majority of existing work focuses on the unpredictable movement and connectivity of the network, making use of techniques like gossip protocols to provide probabilistic assurances of success~\cite{khareWeaving1997,garfinkelPGP1995,eschenauerKeyManagement2002,menesidouCryptographic2017,deandradeFully2016}.
In this paper we argue this is unnecessary:

\begin{hypothesis}
Terrestrial Public Key Infrastructure can be used in space networks, thanks to their predictable links and well-connected segments.
\end{hypothesis}

Instead of designing new protocols, we present backwards compatible modifications to existing terrestrial PKI protocols, ensuring certificate authorities are sufficiently spread out within the network so it is always possible for a device to query an authority within its own segment -- this enables fast connection establishment with low overhead, and the rapid propagation of certificate revocations.
To tackle this problem and support the deployment of secure communication in space systems, we introduce \framework{}, the first experimental framework against which PKI systems can be compared in interplanetary settings, using fixed, well-defined simulation environments in order to ensure consistency.
Using this framework we prove the effectiveness of terrestrial PKI with these modifications, and propose and demonstrate a number of extensions to optimize for the particulars of interplanetary communication, improving performance whilst maintaining compatibility.

\textbf{Contributions.}
In this paper, we establish the goals and requirements for deploying and operating PKI in large satellite networks, with a particular focus on interplanetary networks with long-distance, intermittent relays.
We construct a standardized set of experiments for comparing different key management systems to one another, looking in particular at the performance of connection establishment procedures, and the speed of distributing key revocations and updates in order to minimize the damage caused by a compromised key.
We implement these experiments using the \sysname{} network simulator~\cite{smailesDSNS2025}, extending it to support two commonly used terrestrial PKI systems: Certificate Revocation Lists (CRLs) and the Online Certificate Status Protocol (OCSP).

Our analysis shows for the first time that commonly used terrestrial PKI protocols can be employed in satellite networks.
We also propose and evaluate new configurations within these protocols in order to optimize performance and security in interplanetary contexts, providing more effective revocations, reducing the overhead of connection establishment, and minimizing link congestion.
Our proposals maintain intercompatibility with terrestrial networks, but improve performance under the unique constraints of networking in deep space.

Finally, by providing a standardized experimental framework and implementations in which to test them in simulation, we lay the groundwork for future research, making it easier to empirically demonstrate any improvements over the current state of the art via standardized metrics.
This also benefits network operators, who can easily implement their particular network configuration in simulations to find the optimal configuration within their particular constraints.

Artifacts will be released on publication.

%% file: background.tex
\section{Background}\label{sec:background}

In this section we outline key concepts, including an overview of key management in DTNs and space systems.

\textbf{Delay Tolerant Networks (DTNs).}
DTNs encompass a wide range of networks including mobile ad-hoc networks, wireless sensor networks, and networks in space~\cite{torgersonDelayTolerant2007}.
These networks are characterized by a lack of persistent connectivity and unpredictable network topology.
This requires changes to the network stack, particularly in routing: end-to-end routes cannot always be precomputed, or even established in the first place, due to the structure of the network.
Instead, standards like the Bundle Protocol (BP) are used, enabling store-and-forward delivery~\cite{scottBundle2007}, and proposed routing techniques often rely upon probabilistic methods~\cite{grasicEvolution2011,leguayDTN2005}.\looseness=-1

However, unlike with other types of DTN, the movement of satellites is fixed, so connectivity and optimal routes can be computed ahead of time using techniques like Contact Graph Routing (CGR)~\cite{aranitiContact2015}. However, store-and-forward delivery is still required due to the sporadic nature of links.
Sometimes referred to as the Interplanetary Internet (IPN), a number of works address networking under these particular constraints~\cite{akyildizState2004,elalaouiMARS2020,alhilalSky2019}, including recommendations for architecture and governance from the Internet Society~\cite{ipnsigSolar2023}, and protocol and architecture recommendations from CCSDS~\cite{ccsdsReference2008,ccsdsOverview2014}.

\textbf{Public Key Infrastructure (PKI).}
A PKI attaches identities to public-private keypairs to secure communication in larger networks.
It does this through certificates provided by Certificate Authorities (CAs), which verify a keypair's identity.
A network participant can check a keypair's legitimacy by verifying its certificate, signed by a trusted CA which is part of a signature chain established by the root CA.
This system is near-universal in the terrestrial internet, using the X.509 standard~\cite{boeyenInternet2008}.

On receipt of a signed message, the recipient must check the certificate has not been revoked.
This can be achieved using CRLs, listing all revocations, or using OCSP, which allows devices to query the CA about the status of certificates~\cite{santesson5092013}.
The ``OCSP Stapling'' variant shifts this burden to the sender, who queries the CA and attaches the response to the message, reducing the risk of denial-of-service attacks and the number of handshakes required~\cite{eastlake3rdTransport2011}.
This also improves privacy by preventing certificate authorities from knowing which certificates users are accessing, or which websites they are visiting~\cite{pettersenTransport2013}.

However, in many terrestrial DTNs both CRLs and OCSP are largely dismissed as infeasible~\cite{templinDelay2014,eschenauerKeyManagement2002,djamaludinEstablishing2013}: CRLs are too bandwidth-intensive and OCSP requires constant low-latency access to central CAs, which cannot always be guaranteed.
Instead, DTNs often make use of gossip protocols for key distribution -- nodes exchange information with their peers in the hope that this information spreads as far as possible~\cite{eschenauerKeyManagement2002}.
In this paper we demonstrate that these assumptions do not hold for DTNs in a space context: the predictability of links and the network's topology allow information to be optimally routed, rather than relying upon probabilistic approaches.
We also show that long-distance queries to central CAs can be eliminated by distributing authorities throughout the network, enabling all nodes to make low-latency queries to a local CA present within their network segment.
Once these issues have been addressed, it can be argued that terrestrial PKI is ideal for space applications, scaling particularly well to huge numbers of nodes, thus future-proofing space networks as they continue to grow even larger.

\textbf{Current Systems.}
Many current satellite systems use pre-shared encryption keys~\cite{ccsdsSDLSconcept2024}.
This is significantly simpler to deploy in small systems, as keys can be loaded onto the satellite before launch, but does not scale well to systems with more satellites or shared operation.
It also presents additional risks if keys are leaked, broken, or reverse engineered (see the GK-2A and COMS-1 satellites~\cite{xrit-rx,lrit-key-dec}).
Some systems support over-the-air rekeying in this case, using tightly controlled keys to authenticate the operation of replacing session keys~\cite{andersonAuthentication2023,ccsdsSpace2020}.

The majority of current satellite constellations are privately owned (e.g., Starlink), so less is known about their internal operation.
The Iridium constellation uses pre-shared keys stored in the SIM card, with no mechanism for updating keys short of rewriting or replacing the SIM~\cite{veenemanIridium2021}.
The key management of the satellites themselves is not known.

\section{Related Work}\label{sec:related-work}

Due to its wide range of applications, the issue of security in DTNs and satellite networks has been well-explored, both in scientific literature and in internet drafts and RFCs.

\subsection{Internet Drafts}

The most recent proposed standards for the Bundle Protocol include the BPSec security extensions~\cite{burleighBundle2022,birraneBundle2022}, which add options for integrity and confidentiality to secure BP communication.
Notably, this standard does not recommend a specific method for key management, assuming it to be handled separately~\cite{birraneBundle2022}.
In contrast, in this paper we focus on the key management mechanism, assuming the underlying cryptosystem to be implemented separately.

Other drafts assess the problem of key management in DTNs and space networks.
The authors of~\cite{farrellDelayTolerant2009} recognize key management to be an important open issue in DTNs -- this is expanded on in~\cite{templinDelay2014}, which argues that the long delays and expected disruption in DTNs make traditional PKI methods unsuitable due to their requirement of short-turnaround communication with CAs.

In~\cite{farrellDTN2007}, a basic set of high-level requirements is given for key management in DTNs; they note in particular that systems must account for nodes with highly restricted connectivity, computation, or storage capacity.
The authors of~\cite{templinDTN2016} also set out requirements, noting that the system must not rely on time-constrained queries to a central authority or have a single point of failure, and that revocations must be delay tolerant.

In~\cite{ccsdsSpace2011}, CCSDS gives a high-level overview of many potential key management concepts in space DTNs.
There is also an ongoing CCSDS project to design a cloud-based CA for satellite systems, to be used as a common route of trust particularly in shared systems~\cite{igcs_ccsds_357}.
Results from our simulations will help to inform this and other projects -- this is discussed further in Section~\ref{sec:discussion}.

In~\cite{viswanathanArchitecture2016}, the authors lay out five potential design patterns for PKI in DTNs, which we use as one basis for our simulations.
The ``request-response'' pattern describes OCSP and OCSP Stapling, ``publish with proxy subscribe'' matches behavior seen in OCSP with Validators, and ``blacklist broadcast'' resembles CRLs.

\subsection{Academic Works}

Menesidou et al.~\cite{menesidouCryptographic2017} provide an overview of works looking at key management in DTNs.
Some of these focus on space networks or are network-agnostic, but the majority look at less relevant applications of DTNs such as rural networks.

A number of works focus on security initialization or authenticated key exchange -- that is, initiating a secure channel where one did not previously exist by exchanging or establishing keys~\cite{menesidouAuthenticated2012}.
We consider these to be out of scope of this paper, and focus on communication and updates over an already established network.

The authors of~\cite{bhuttaSecurity2009} recognize that terrestrial protocols do not perform well in DTNs and smaller satellite networks, due to their reliance on constant connectivity and low latency communication.
They also acknowledge that any key management system will need to account for the types of nodes and heterogeneity in the network.

In~\cite{vonmaurichData2018} the authors focus on real-world assessments of cryptosystems on embedded hardware.
Additionally, they discuss the requirements of security protocols in federated satellite systems.
They propose a high-level PKI-based protocol to guarantee confidentiality, integrity, and authenticity, and rely on a centralized authority to verify and distribute keys.

Some works such as~\cite{johnsonProviding2013} adapt PKI for use in satellite networks -- their system is very similar to OCSP stapling systems, with nodes requesting an authentication pass from a CA ahead of communication.
On the other hand, \cite{sethPractical2005} argues that PKI is not suitable for use in DTNs, instead suggesting the use of Hierarchical Identity Based Cryptography (HIBC), in which keys are derived from public information about nodes, and a hierarchy of authorities manages keys.
This is argued to be better than traditional PKI, as the public parameters do not need to be verified as frequently~\cite{asokanSecuring2007}, at the cost of requiring distribution of additional secrets out of channel.
We argue that this is not necessary for satellite networks, as each network segment is sufficiently connected that it is always possible to query a trusted authority, provided there are sufficiently many distributed throughout the network.

The authors of~\cite{deandradeFully2016} propose a decentralized key management system by passing around chains of certificates, with nodes added to the chain when another node trusts it.
While fully decentralized, it has has no mechanisms for key revocation or centralized/federated management.

In~\cite{koisserVCER2022}, Koisser et al.\ propose a gossiping protocol based on sparse Merkle trees to support PKI in constrained settings such as satellite networks.
They demonstrated their approach using an abstracted network simulation and on the OPS-SAT satellite, a computationally powerful satellite which was used for experiments by industry and academia.
However, Merkle trees are more computationally intensive than traditional methods, and may be difficult to run on more constrained satellite hardware.
Furthermore, their approach is not compatible with terrestrial PKI, so a compatibility layer would be required for communication with terrestrial networks.

Finally, several works extend existing PKI concepts towards use in DTNs.
The authors of~\cite{cooperMore2000} and~\cite{bhuttaPublic2017} build upon traditional CRLs, reducing message size and storage by sending ``delta CRLs'' containing incremental updates and by using hash tables.
In~\cite{koisserTruSat2022}, the authors build a consensus protocol to remove the need for a single root CA, and test a number of revocation methods including OCSP Stapling and Bloom filters.
The authors of~\cite{luoDICTATE2005} recognize that a single centralized CA is not enough for PKI in DTNs, and propose the use of distributed CAs to ensure reachability.
In~\cite{fangAdaptive2013} this concept is extended to ``adaptive CAs'', in which any node can become a CA, and decisions are made dynamically.

%% file: threat-model.tex
\section{Threat Model}\label{sec:threat-model}

In this paper we consider an attacker that has gained access to the private key of a legitimate network user, and has at least one node in the network under their control (e.g., a satellite).
Using this key they can craft messages as if they were the legitimate user.
This could occur by compromising a satellite (e.g., by compromising telecommand authentication, taking control of a satellite via a compromised ground segment, or via a malicious hosted payload), intercepting improperly encrypted messages, or compromising the supply chain.

The high-level goal of the attacker is to disrupt operation within the network, by using a compromised key to communicate with as many nodes as possible, for as long as possible.
The specifics will look different depending on the network and the compromised device: for example, by producing forged measurement data, impersonating users in communication, or forging telecommand packets to cause physical damage to the device.

Once the attacker is discovered, the compromised key must be revoked.
Since there are long distances (and therefore high latencies) between nodes in an interplanetary network, the goal is to minimize the extent of damage from the moment the attack is identified.
This is achieved by propagating revocation data as efficiently as possible.
To persist for as long as possible in the network, the attacker may attempt to jam the links most critical to the revocation, such as relay links, or links adjacent to a certificate authority.

%% file: framework-design.tex
\section{Design of \framework{} Framework}\label{sec:framework-design}

We now establish the goals and requirements for efficient key management in interplanetary networks, and use them to construct the \framework{} framework: a set of standardized experiments and derived metrics, which can be used to compare protocols against one another in a consistent environment using simulations.
\framework{} is neither limited to a specific network topology nor a particular PKI protocol.
This enables researchers and standards bodies to test protocols and configurations against one another in a standardized, repeatable manner.
Alongside supporting research, \framework{} also enables satellite operators to implement the particulars of their network's topology and optimize a protocol for their exact configuration, thus providing data-based decision making about which configuration to use.

We state the following goals for PKI in satellite networks (later formalized as standard metrics):
\begin{enumerate}[label=\textbf{G\arabic*:}, ref=\textbf{G\arabic*}, leftmargin=*]
    \item\label{itm:latency} \textbf{Low latency.} The latency of messages should be as close as possible to the shortest path between the two nodes, minimizing the additional distance traversed during a connection handshake.
    
    \item\label{itm:establishment-overhead} \textbf{Low establishment overhead.} This is the time difference between a full connection handshake between two nodes and the time taken to send a single message. It is crucial that this is minimized due to the high latency of relay links and potentially short contact windows.
    
    \item\label{itm:revocation-coverage} \textbf{Fast revocation coverage.} When a certificate is revoked, information about that revocation must be propagated to all the other nodes in the network as quickly as possible.
    
    \item\label{itm:attack-resilient} \textbf{Resilient against jamming attacks.} The PKI mechanism must maintain performance even under adverse interference, with attackers jamming specific links.
    
    \item\label{itm:link-saturation} \textbf{Low link saturation.} Inter-satellite links are often highly bandwidth constrained. Systems should therefore avoid sending too much traffic via a single node to avoid overwhelming their capacity.
    
\end{enumerate}

We also suggest the following as desirable secondary goals:
\begin{enumerate}[label=\textbf{S\arabic*:}, ref=\textbf{S\arabic*}, leftmargin=*]
    \item\label{itm:distributed} \textbf{Distributed.} In order to achieve good performance across sporadic high-latency links, authorities should be distributed across network segments, enabling querying without excessive latency. We discuss in Section~\ref{sec:discussion} the trade-off between centralized control and performance.
    \item\label{itm:intercompatible} \textbf{Intercompatible.} By working alongside existing internet infrastructure, systems are more likely to be integrated into an interplanetary internet. This may be achieved through direct compatibility or a translation layer.
    \item\label{itm:computation} \textbf{Efficient computation.} Computation and validation of certificates should not require excessive computational power, in order to work within the constrained hardware requirements of many satellite systems.
\end{enumerate}

\subsection{Scenarios and Metrics}

We evaluate how well a PKI system achieves these goals through two imperative scenarios: connection establishment and key revocation.
From these scenarios we extract aggregated statistics (e.g., latency, link saturation) to compute scores regarding each of the core goals.
We extract these scores by taking the mean, in addition to looking at the distributions directly to gain further insights.

\subsubsection{Connection Establishment}

In the first scenario we consider connection establishment: two nodes wish to communicate, but the sender must first prove its identity to the receiver by sending a signed message with its certificate attached, validated by the recipient (typically by requesting information from a certificate authority).
This scenario is designed to test goals \ref{itm:latency}, \ref{itm:establishment-overhead}, and \ref{itm:link-saturation}.

To eliminate traffic modeling as a confounding variable we consider connection establishment between every pair of communicating nodes in the network, giving a distribution of values.
These are aggregated into scores across the whole network, and in interplanetary networks may optionally be divided into distinct segments -- for instance, into groups of nodes belonging to Earth, the Moon, or Mars. We use the following metrics:

\paragraph{Delivery Metric.}
Proportion of delivered messages.

\paragraph{Latency Metric.}
Mean of the overall latency of delivered messages.

\paragraph{Overhead Metric.}
The overhead introduced by the PKI system -- for each message, this is calculated by taking the overall latency of the message exchange (including any PKI/verification mechanisms) and subtracting the latency of the original message. The mean is then taken across all messages.

\paragraph{Saturation Metric.}
The severity of full link saturation across the network.
Computed by finding the amount of time a link is fully saturated during the simulation, summed across all links.

\subsubsection{Key Revocation}

In the second scenario we consider the process of revoking a key in use by a malicious actor.
Once a key has been revoked, the system should ensure the revocation propagates across the network as quickly as possible, and minimize the reach of the attacker past this point.
This scenario tests goals \ref{itm:revocation-coverage} and \ref{itm:attack-resilient}.

We again compute metrics per-segment, although this time we need multiple segments for a full understanding of performance, since revocations will propagate to different network segments at different speeds.
This often causes a race between messages and thus substantially different results between segments (see Appendix~\ref{app:revocation-coverage}).
We use two coverage metrics:

\paragraph{Coverage Metric.}
This metric is computed to be the time at which all nodes in a victim segment $S_V$ are protected from malicious messages sent by an attacker in segment $S_A$, when the revocation is issued by a CA present in segment $S_R$.
This value may be negative, indicating the revocation has fully propagated across the victim segment before any messages from $S_A$ have the chance to reach $S_V$.
To reduce the number of values to consider, the mean and maximum values are taken across all triples of segments $(S_V, S_A, S_R)$.

\paragraph{Attacked Coverage Metrics.}
To assess resilience against jamming attacks, coverage is measured when loss is induced on specific links following revocation: the links surrounding the CA, and the relay links between segments.
The relay links are easy for an attacker to access due to their proximity to the ground, and are likely to cause significant impact if disrupted.
However, they are very high-power links and will require similarly powerful hardware to successfully jam.
The CA links are comparatively low-power, matching those of the surrounding nodes, but are harder to reach due to being in orbit.
If an attacker can disrupt these links, they will be able to block revocation messages from reaching their target.

On the targeted links, loss is fixed at $p=0.1$ (``weakly attacked'') and $p=0.5$ (``strongly attacked'').
Existing work has shown that the CCSDS-standard telecommand packets can be disrupted by an attacker with a jamming-to-signal ratio between \num{-8.93}~and \qty{1.00}{\decibel}, and that a \qty{50}{\percent} error rate can be caused at \qty{-11.98}{\decibel}~\cite{salkieldSpaceJam2025}.
These results vary quite widely and the hardware available to an attacker can range from cheap off-the-shelf components through to expensive high power amplifiers, but it is clear that a range of attacks are possible.
We consider two classes of jamming-capable attacker to sufficiently cover most attack cases; should operators wish to evaluate other attacks, the simulation scenario is easy to modify and extend to cover additional cases.

Coverage metrics are computed and reduced in the same way as before.
Note that all messages eventually reach their destination due to the guaranteed delivery provided by LTP, but lost messages result in retransmits, increasing latency.
These messages may in turn be lost, but all messages will eventually be delivered.

\subsection{Further Considerations}

For the sake of reproducibility, we also fix the following:

\subsubsection{Parameter Selection}

To ensure experiments remain consistent across differing implementations, we must ensure values like message size, maximum segment size, and maximum retransmission count are well-defined. These parameters are stated, and the choice of values explained, in Appendix~\ref{app:parameter-selection}.

\subsubsection{Routing}\label{sec:framework-design-routing}

Routing in DTNs and satellite networks is a difficult problem, and has been explored extensively in recent literature~\cite{grasicEvolution2011,leguayDTN2005,aranitiContact2015,sobinSurvey2016,xiaogangSurvey2016,alagozExploring2007}.
Difficulties stem from constantly changing connectivity and many-hop paths, requiring routing that can react to changes more rapidly than update messages can traverse the network.
However, the topology of many interplanetary networks makes it easier to route traffic -- the use of only a small number of relays between segments means traffic only needs to be routed within each segment.
Messages bound for other planets simply need to route to the relay, which can handle routing for the remainder of the journey.

To prevent biasing results by a particular routing strategy, all nodes have access to optimal routing information at all times, including looking ahead to the future state of the network in store-and-forward routing.
This provides results equivalent to source routing, in which a message's route is computed at the source and attached to the message, so it can simply be forwarded along the computed path.
This approach is ideal for resource-constrained settings, since there is no need for intermediate nodes to compute or store routing tables.
Handling routing in this way ensures we eliminate the choice of routing protocol from impacting our results -- however, should future work wish to compare routing protocols, this can easily be done by swapping out the routing component and leaving the simulations otherwise unchanged.

\subsubsection{Traffic Modeling}

We do not rely upon randomized traffic modeling, and instead use pairwise communication between every pair of ground nodes in the network. This provides us with a distribution for each performance metric, which can be further partitioned into network segments, enabling a comprehensive analysis of the network's performance under each PKI configuration.

%% file: experiment-design.tex
\section{Experiment Design}\label{sec:experiment-design}

As part of \framework{} we provide a set of standard simulations, following the framework established in the previous section.
These simulations have been built on top of the ``\sysname{}'' simulator due to its ability to scale to many nodes and support of long-distance links, enabling us to easily construct the scenarios defined in the previous section~\cite{smailesDSNS2025}.
The simulator already has built-in support for reliable message routing and delivery, and we extended it by implementing each of the PKI mechanisms evaluated in this paper.
We were also able to extract all the desired metrics by building upon the simulator's basic logging functionality, and to build each of the network topologies described in the previous section by extending the reference scenarios provided with the simulator.
Our contributions will be released as an extension to \sysname{} on publication, enabling future work to implement additional PKI mechanisms and compare directly to our results.

\subsection{Network Topologies}

We now design a set of representative network topologies against which we test the \framework{} framework. These topologies cover common use cases for real-world networks and serve as a starting point if operators desire to implement their own networks.
We implement these by extending the custom scenarios provided by the \sysname{} simulator~\cite{smailesDSNS2025}.
The base latency of communication between two points in each network topology is shown in Appendix~\ref{app:latency-results}.

\subsubsection{Earth, Moon and Mars Network}

\begin{figure}
    \centering\includegraphics[width=.9\linewidth]{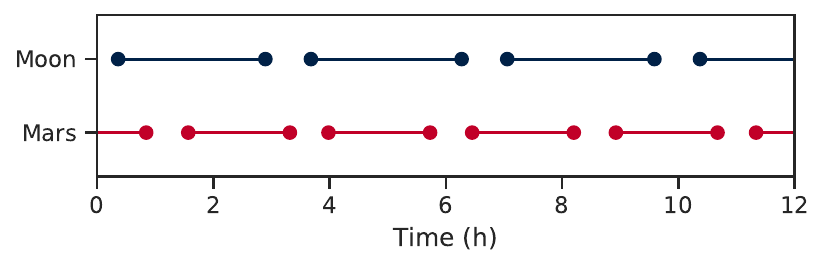}
    \caption{State of the relays from Earth to the Moon and Mars over time. Gaps indicate that the link is not available, resulting in a split between network segments.}
    \label{fig:segments-earth-moon-mars}
\end{figure}

Since our primary focus is on communication in interplanetary networks, we focus on a representative network with communication between the Earth, the Moon, and Mars. We match existing proposals for future interplanetary networks, such as NASA's ``LunaNet'' proposal, in which relays enable communication between nodes on/orbiting Earth and the Moon~\cite{israelLunaNet2020}. The network contains \num{66}~satellites matching the Iridium constellation, and \num{256} randomly placed ground stations (see Appendix~\ref{app:experiment-configuration}).\footnote{Each ground station has a minimum elevation angle of \qty{8.2}{\degree}, matching Iridium specifications~\cite{prattOperational1999}.}
Mars has \num{66} satellites and \num{12} ground stations, and the Moon has \num{8} satellites and \num{12} ground stations. To relay data between segments, \num{3}~ground stations matching NASA's Deep Space Network are used, connected to relay satellites around the Moon and Mars, with orbital periods of approximately \num{3.3}~hours and \num{2.5}~hours respectively.\footnote{The Moon relay has an orbital period of~\qty{11941}{\second} and the Mars relay has an orbital period of~\qty{8829}{\second}.} Although near-constant communication would be possible by adding more relay satellites, we are interested in assessing whether tested PKI systems can maintain functionality even when connectivity between segments is interrupted.

Due to the change in the state of relay links over time (illustrated in Figure~\ref{fig:segments-earth-moon-mars}), this network topology must be evaluated at \num{4} different start times to cover every possible combination of relay states (both connected, both disconnected, only one connected).
Results are averaged across all \num{4} relay states.

\subsubsection{Earth Constellation}

We evaluate a single-constellation network as a control, to compare the properties of a well-connected network against interplanetary networks. To match a real-world network, we use a Walker constellation with \num{66}~satellites, like the Iridium constellation, again connected to \num{256}~ground stations.

\subsubsection{Federation of CubeSats}

Finally, we take the ``CubeSat Constellation'' scenario provided by \sysname{}, in which a federated network is constructed from the \num{98}~CubeSats currently in orbit.\footnote{Positions are derived from Two-Line Element sets provided by \href{https://celestrak.org/}{celestrak.org} on 2025-06-27 and propagated from 2025-06-27T00:00:00Z. Up-to-date TLEs may be used, but results will not be identical.}
CubeSats are connected to the same \num{256}~ground stations as before, and to each other using inter-satellite links with a range of \qty{2500}{\kilo\metre}.
The dynamic mesh formed by this network has very short contact windows, providing an interesting challenge for PKI: the entire session must be negotiated, including certificate validity checks, during the brief period during which a connection is available.

\subsection{Key Management Systems}

\begin{figure*}
    \centering\includegraphics[width=.9\textwidth]{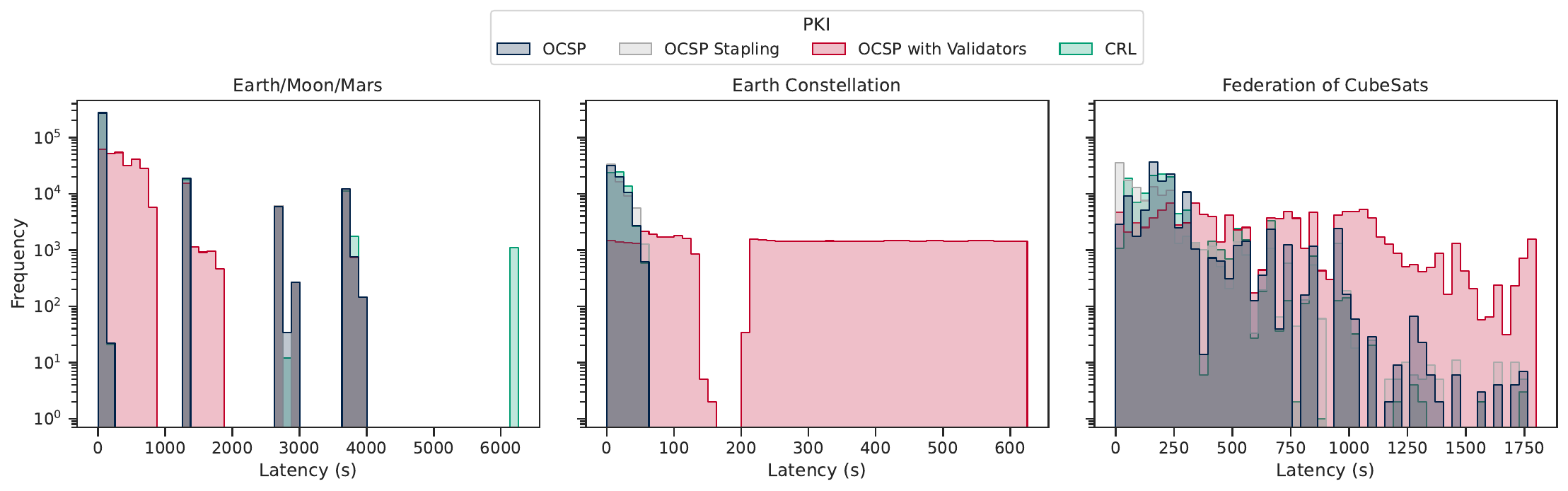}
    \caption{Latency distribution when establishing a new connection under each network topology, for each base PKI configuration with a single centralized CA.}
    \label{fig:establishment-latency-constellations}
\end{figure*}

\begin{table}
    \caption{Full results for the connection establishment scenario for each of the base PKI configurations. The best and worst results are highlighted in blue and red, respectively. EMM = Earth, Moon, and Mars; E = Earth Constellation; C = Federation of CubeSats.}
    \label{tab:results-establishment-constellation}
    \autobox{
    {\sisetup{round-mode=figures,round-precision=4}
    \begin{tabular}{ccccS[table-format=3.2]S[table-format=3.2]S[table-format=3.3]S[table-format=5.1]}
        \toprule
        & \multicolumn{3}{c}{Configuration} & \multicolumn{4}{c}{Metrics} \\
        \cmidrule(lr){2-4} \cmidrule(lr){5-8}
        Network & Dist. & PKI & Variant & {Delivered (\%)} & {Latency} & {Overhead} & {Saturation} \\
        \midrule
        EMM &  & CRL & -- & 100.00 & 412.52 & 118.36 & 993.25 \\
        EMM &  & OCSP & -- & 100.00 & 408.23 & 115.75 & 988.75 \\
        EMM &  & OCSP & Stapling & 100.00 & 410.53 & \worst{211.00} & 768.50 \\
        EMM &  & OCSP & Validators & 100.00 & \worst{604.70} & 67.52 & 2411.50 \\
        EMM & \cmark & CRL & -- & 99.82 & 318.42 & 24.26 & 991.25 \\
        EMM & \cmark & OCSP & -- & 100.00 & 299.01 & \best{4.85} & 986.75 \\
        EMM & \cmark & OCSP & Stapling & 100.00 & \best{298.36} & 7.12 & \best{737.25} \\
        EMM & \cmark & OCSP & Validators & 100.00 & 579.55 & 48.73 & \worst{2995.33} \\
        \midrule
        E & -- & CRL & -- & 100.00 & 17.92 & 7.84 & 1386.00 \\
        E & -- & OCSP & -- & 100.00 & \best{15.86} & \best{5.78} & 1391.00 \\
        E & -- & OCSP & Stapling & 100.00 & 16.91 & 9.60 & \best{925.00} \\
        E & -- & OCSP & Validators & 100.00 & \worst{325.83} & \worst{45.84} & \worst{3488.00} \\
        \midrule
        C &  & CRL & -- & 99.72 & 196.14 & \best{77.90} & 6525.00 \\
        C &  & OCSP & -- & 100.00 & 249.08 & 129.15 & 6559.00 \\
        C &  & OCSP & Stapling & 100.00 & 160.26 & 86.72 & 5335.00 \\
        C &  & OCSP & Validators & 100.00 & \worst{792.71} & \worst{217.37} & 14283.00 \\
        C & \cmark & CRL & -- & 94.35 & 205.59 & 87.31 & 6538.00 \\
        C & \cmark & OCSP & -- & 100.00 & 236.06 & 116.12 & 6553.00 \\
        C & \cmark & OCSP & Stapling & 100.00 & \best{144.26} & 78.39 & \best{4683.00} \\
        C & \cmark & OCSP & Validators & 100.00 & 630.77 & 159.44 & \worst{14637.00} \\
        \bottomrule
    \end{tabular}
    }}
\end{table}

We implement a number of PKI systems based on the existing and proposed mechanisms discussed in Section~\ref{sec:related-work}.
In each case, nodes have a public/private keypair which is used to sign and encrypt messages.
These keypairs are accompanied by a certificate -- that is, a chain of signatures tracing back to a Certificate Authority (CA) vouching for its legitimacy.
We implement the following systems to verify a certificate's validity, ensuring it has not been revoked:
\begin{itemize}
    \item \textbf{Certificate Revocation Lists (CRLs):} CAs each maintain a list of revoked certificates, which are sent to nodes on request. Any certificates on this list are rejected, and others are accepted provided there is a trusted CA in the certificate chain. CRLs are cached, with a new CRL requested on the next message received after expiry. To reduce size, CRLs can be built using hash tables~\cite{bhuttaPublic2017} or using delta CRLs~\cite{cooperMore2000}.
    \item \textbf{Online Certificate Status Protocol (OCSP):} On receipt of a signed message, nodes request the certificate's validity information from their CA. The response is cached to reduce the overhead on subsequent messages.
    \item \textbf{OCSP Stapling:} Similar to the above, but the node sending the message requests its CA for a signed message with validity information. This is sent alongside the original message, reducing the burden on the recipient.
    \item \textbf{OCSP with Validators:} Proposed in~\cite{viswanathanArchitecture2016}, messages are instead sent \textit{via} a CA, which staples the certificate validity to the message in-transit. This removes the need for a full handshake in certificate validation, and can significantly reduce latency if the CA is en route to the message destination. This shares some similarities with the proposed ``Revocation in the Middle'' technique, using middleboxes to provide revocation information~\cite{szalachowskiRITM2016}.
\end{itemize}

We also propose and evaluate a number of new techniques which alter the behavior of OCSP in order to provide improvements specific to the topology and other constraints of interplanetary networks.
These are as follows:
\begin{itemize}
    \item \textbf{OCSP Hybrid:} We note that in-transit OCSP validation increases congestion in a lot of cases by sending all traffic via the CA, but makes a lot of sense when traffic is traveling via a small set of relays anyway (e.g., traffic between Earth and Mars). We therefore propose a hybrid system in which standard OCSP Stapling is used for traffic within a segment, but any cross-segment traffic uses in-transit validation, with CAs located at the relays between network segments.
    \item \textbf{OCSP Parallel:} Although using in-transit validation can reduce the overhead of establishing a connection, it risks congesting the links closest to the CAs, since a large number of messages traverse these links. We propose instead that only the certificate validity information is sent via the CA, with the remainder of the message sent directly, thereby gaining the benefits of both lower latency and reduced congestion. This is only used for messages within a segment, since cross-segment messages will be sent via the relay-based CA regardless.
    \item \textbf{Relay Firewall:} In interplanetary networks, cross-segment traffic must travel via relays -- this presents an excellent opportunity to remove messages from revoked keys before reaching their destination. When the relay is informed about a revocation, it scans incoming messages for the revoked certificate and drops any that match, in addition to its usual CA functionality. This can significantly reduce load on the relay links by dropping attackers' messages much sooner, and has the potential to reduce the number of nodes reachable by an attacker with a revoked key.
\end{itemize}

\begin{figure*}
    \centering\includegraphics[width=.9\textwidth]{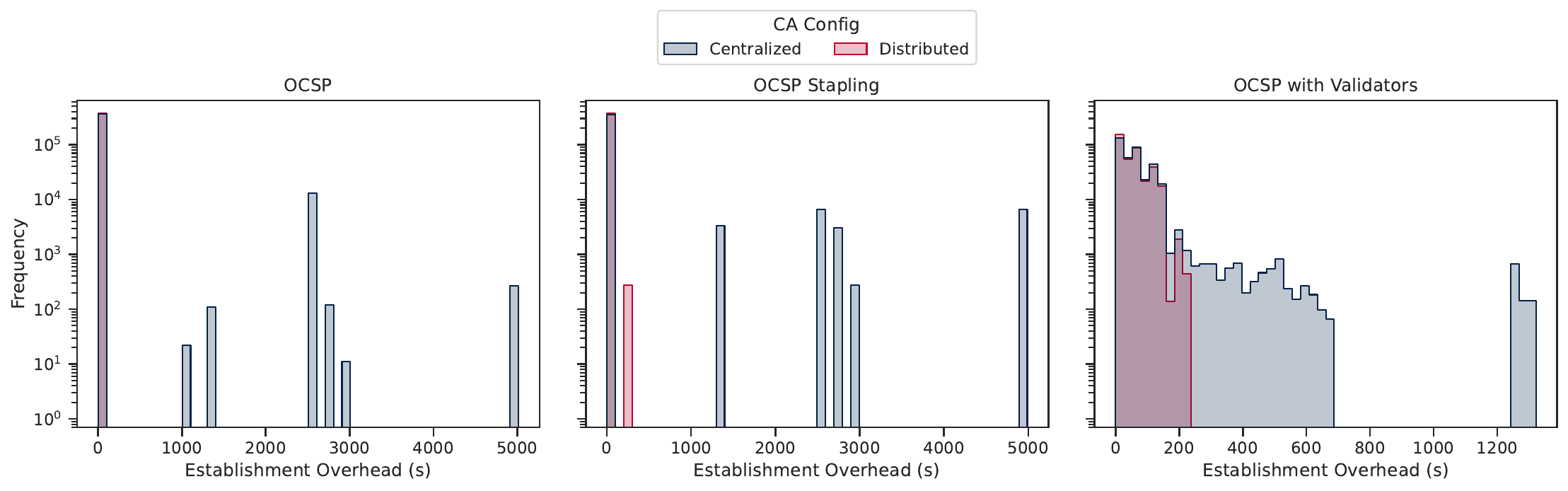}
    \caption{Establishment overhead of each of the base OCSP configurations on the Earth/Moon/Mars network, with centralized and distributed authority. Overhead is significantly higher in the centralized cases since all traffic must travel via Earth, resulting in very long round-trip times.}
    \label{fig:establishment-overhead-centralized-distributed}
\end{figure*}

\begin{figure*}
    \centering\includegraphics[width=\textwidth]{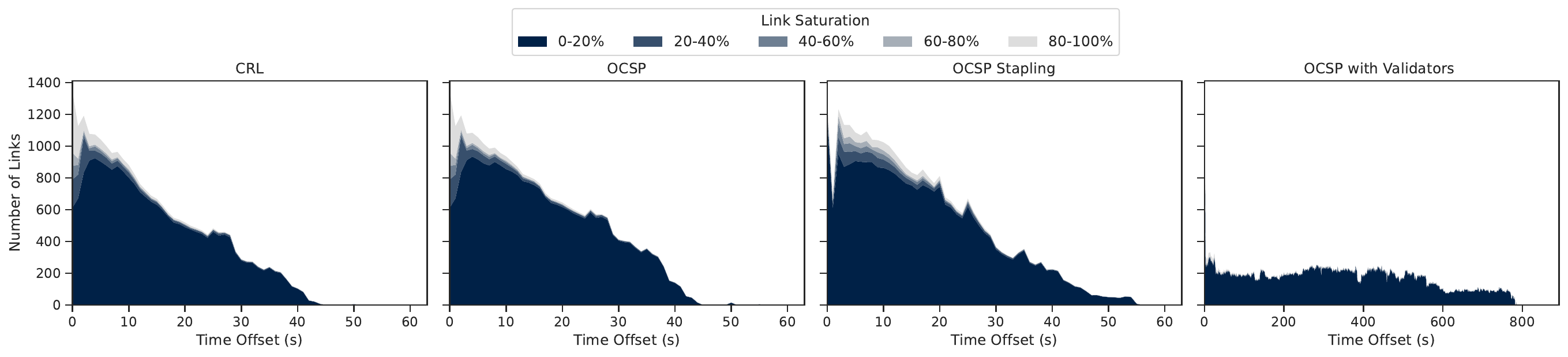}
    \caption{Link saturation over time for each base PKI configuration. Note that OCSP with Validators uses an extended time axis, since the message queues on some links take much longer to resolve -- sending all traffic via CAs results in a huge number of messages queued over a small number of routes.}
    \label{fig:saturation}
\end{figure*}

Sequence diagrams for each of these techniques, in addition to the base CRL/OCSP configurations, can be found in Appendix~\ref{app:protocols}.
We evaluate each of these under a ``centralized'' model, with a single terrestrial CA, and a ``distributed'' model.
In the distributed case, there is one CA per network segment (i.e., layer of satellites or ground systems), which are designated as sub-CAs of a single terrestrial root CA, such that any CA can provide validity information for any other certificate.
When a certificate is issued, updated, or revoked under this model, the central CA sends a signed message with the identity of the changed certificate to each of the sub-CAs.
In the case of revocation, sub-CAs then reject validity checks for that certificate received after the update.
This approach can significantly reduce the latency overhead of validating cross-segment messages by providing low-latency access to sub-CAs, but can also increase the threat of attackers exploiting revoked certificates from distant segments due to slow updates.
We later demonstrate how this can be prevented through the inclusion of a relay firewall.

\subsection{Additional Configuration}

As previously discussed, \framework{} does not specify a particular routing protocol, instead assuming all nodes have access to optimal routing information at all times.
This is achieved using the store-and-forward routing provided by \sysname{}, which looks forward to the state of the network at fixed timesteps in order to compute routes that involve temporary storage of messages.

The experiments have been constructed so that cache duration does not need to be considered, since OCSP and CRL responses do not need to be cached to measure first-time establishment overhead or the coverage of a revocation.
In real usage, responses from the CA will be cached for a specified amount of time to reduce the need for redundant requests when communication between two nodes is frequent.
Our implementations of the PKI protocols support configurable cache times, and operators can use our simulations to optimize cache times for the particulars of their network topology once the broad protocol architecture has been decided.

\subsubsection{Message Delivery}

Message routing and delivery is handled by the simulator.
In addition to standard message routing, we also use \sysname{}'s implementation of Licklider Transmission Protocol (LTP).
This protocol is designed for deep space links, used as a convergence layer protocol underneath Bundle Protocol or other higher layer protocols~\cite{burleighLicklider2008}.
LTP provides guaranteed delivery by splitting messages into segments, and reporting any failed deliveries.
This ensures messages are eventually delivered in the attack scenarios, albeit with higher latency.

\subsubsection{Log Processing}\label{sec:log-processing}

\begin{table*}
\footnotesize
    \caption{Time taken for revocation to cover all nodes for the the Earth/Moon/Mars constellation with a distributed CA. Results are aggregated by attacker segment, revocation origin, and victim segment. Best/worst results for each attack type highlighted.}
    \label{tab:results-revocation-interplanetary}
    \autobox{
    {\sisetup{round-mode=figures,round-precision=4}
    \begin{tabular}{cccS[table-format=-4.1]S[table-format=4.0]S[table-format=-4.1]S[table-format=4.0]S[table-format=-4.2]S[table-format=4.0]}
        \toprule
        & & & \multicolumn{6}{c}{Coverage Time (s)} \\
        \cmidrule(lr){4-9}
        \multicolumn{3}{c}{Configuration} & \multicolumn{2}{c}{No Attack} & \multicolumn{2}{c}{Weakly Attacked ($p=0.1$)} & \multicolumn{2}{c}{Strongly Attacked ($p=0.5$)} \\
        \cmidrule(lr){1-3} \cmidrule(lr){4-5} \cmidrule(lr){6-7} \cmidrule(lr){8-9}
        PKI & Variant & Attack & {Mean} & {Max} & {Mean} & {Max} & {Mean} & {Max} \\
        \midrule
        CRL & -- & CA & \best{-104.45} & \worst{1438.80} & \best{-104.22} & 1439.03 & -99.53 & 1446.39 \\
        OCSP & -- & CA & \best{-104.45} & \worst{1438.80} & \best{-104.22} & 1439.03 & \best{-101.21} & 1443.16 \\
        OCSP & Stapling & CA & \worst{1366.11} & \best{1409.38} & 1366.26 & \best{1409.54} & 1369.09 & 1413.91 \\
        OCSP & Validators & CA & \worst{1366.16} & \best{1409.41} & 1366.46 & \best{1409.79} & 1368.31 & \best{1412.08} \\
        \midrule
        CRL & -- & Relay & \best{-104.45} & \worst{1438.80} & 346.80 & \worst{2368.39} & 2670.41 & \worst{6460.43} \\
        OCSP & -- & Relay & \best{-104.45} & \worst{1438.80} & 346.80 & \worst{2368.39} & 2670.41 & \worst{6460.43} \\
        OCSP & Stapling & Relay & \worst{1366.11} & \best{1409.38} & \worst{1671.80} & 2016.18 & \worst{4081.13} & 5722.06 \\
        OCSP & Validators & Relay & \worst{1366.16} & \best{1409.41} & \worst{1671.83} & 2016.20 & \worst{4081.14} & 5722.06 \\
        \bottomrule
    \end{tabular}
    }}
\end{table*}

\begin{figure*}
    \centering\includegraphics[width=.9\textwidth]{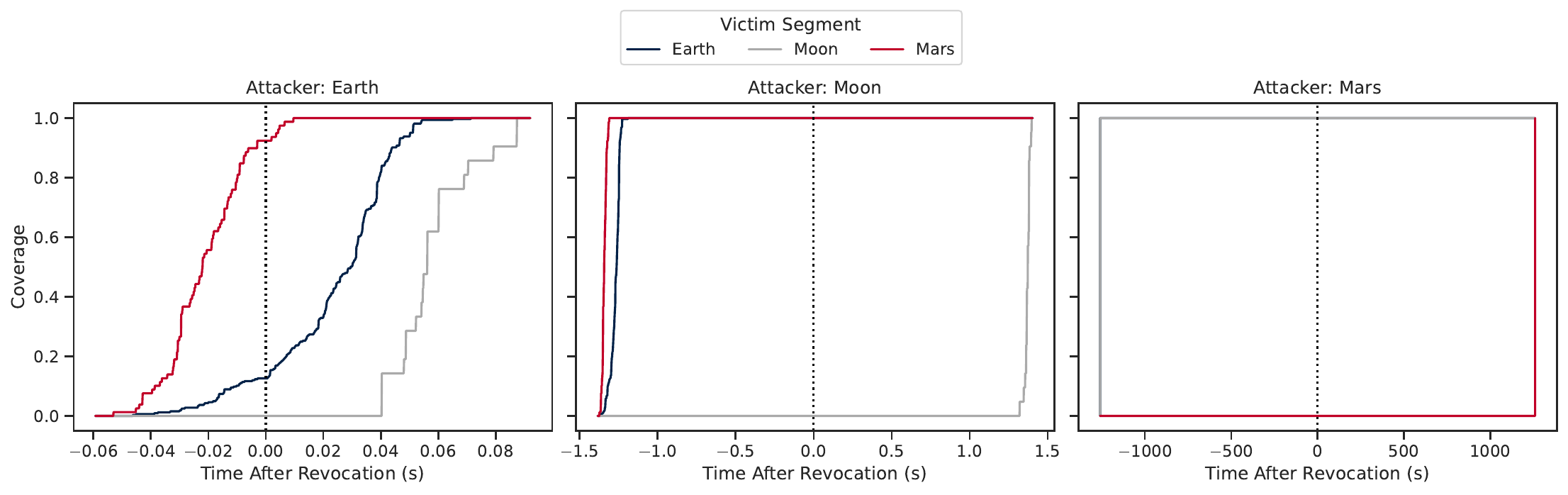}
    \caption{Coverage over time for a revocation, relative to the time at which the revocation was issued. Revocation originates from an Earth-based CA, under the base OCSP configuration with distributed CAs.}
    \label{fig:coverage-interplanetary}
\end{figure*}

The output of a \sysname{} simulation is a log of all events during the simulation.
Due to the large number of events and the limited usefulness of single events, we first group these by message, turning a series of message send/receive events into a single logging object with latency, hop count, etc.
We then group messages together to associate CA queries with the underlying message for which they were created, to extract more useful statistics (e.g., latency of the exchange, establishment overhead).
The specifics are described in Appendix~\ref{app:data-processing}.
We also use a separate logging actor to measure link saturation, which looks at how per-link queue lengths change over time.

%% file: results.tex
\section{Results}\label{sec:results}

We now evaluate each of the proposed PKI techniques using \framework{}. We start by looking at baseline results for the establishment and revocation scenarios, followed by assessing each of our newly proposed techniques to see if they provide a measurable improvement upon the baseline.

\subsection{Connection Establishment}\label{sec:results-establishment}

First, we examine the connection establishment cost under each configuration under each of the three network topologies.

The latency distribution under a selection of centralized configurations can be seen in Figure~\ref{fig:establishment-latency-constellations}, and full results are given in Table~\ref{tab:results-establishment-constellation}.
Figure~\ref{fig:establishment-overhead-centralized-distributed} further demonstrates the establishment overhead for the Earth/Moon/Mars network topology, comparing the centralized and distributed configurations.

From these we can gain some interesting insights about the effectiveness of communication in each constellation, and of each PKI configuration.
It is immediately clear that centralized PKI is infeasible in interplanetary networks, due to the requirement that each connection establishment communicates with the Earth-based CA: as expected, the overhead of making these centralized connections is simply too high, and a large number of messages are dropped.
However, when authority is properly distributed, even in the interplanetary setting, performance under both OCSP and CRL configurations is significantly improved -- establishment overhead, latency, and link saturation are all significantly reduced, with results comparable to those of the Earth-based constellations.

We also see that the CubeSat network has high latency, and struggles to route all messages -- this is likely due to the satellites' limited coverage and intermittent connectivity.
If simulations run for long enough it will likely eventually route messages, so such a network may be suitable for certain use cases, but making calls back to a centralized CA will incur significant latency costs.

Finally, we can see that the ``OCSP with Validators'' configuration struggles with high establishment overhead, even in distributed cases, despite the fact that this configuration removes the need for a round-trip handshake with the CA.
The root cause of this issue is that all messages are routed via a small number of CA nodes, causing the links surrounding them to become immensely congested.
This can be seen in Figure~\ref{fig:saturation}, in which it takes over \num{10}~times longer for links to return to normal following the start of communication.
We explore in Section~\ref{sec:results-improvements} how our proposed changes to OCSP can improve performance in this area.

\subsection{Key Revocation}\label{sec:results-revocation}

We examine the revocation scenario to see how well each PKI mechanism handles revoked keys, and how quickly they can distribute information about the revocation across the network. We focus on the Earth/Moon/Mars scenario.

Figure~\ref{fig:coverage-interplanetary} illustrates the coverage time for a revocation under the base OCSP configuration.
We can see that the coverage time is often negative for segments in which the attacker is not present -- as discussed in Section~\ref{sec:framework-design}, this is due to the long distances between segments, enabling the revocation to reach the CA long before the attacker's message traverses the relay link.
As a result, only messages sent significantly before of the revocation will be accepted in that segment, resulting in a negative coverage time.

In contrast, when the attacker and revoker are located in the same segment, coverage time is primarily influenced by each node's proximity to the CA, leading to rapid overall coverage with some nodes protected later than others.
Through proper configuration, we aim to achieve the earliest possible coverage in each scenario, minimizing damage within the speed-of-light constraints inherent in long-distance communication.

Aggregated coverage information for each of the distributed configurations is given in Table~\ref{tab:results-revocation-interplanetary} (centralized results are available in the full dataset).
The overall coverage statistics are quite similar between configurations, but there are some differences: CRLs and base OCSP have better mean coverage than either of the other OCSP configurations.
This is because the recipient must query the CA, rather than the sender, giving more time for the revocation to propagate before the attacker's message reaches its destination.
Worst-case performance remains fairly consistent between configurations since it is primarily governed by the speed-of-light delays between planetary segments.
Therefore, our proposed improvements to the protocols focus on reducing the mean coverage time.

Finally, we can see that attacking the relays is far more successful than attacking the links surrounding the CA, but that broad performance trends remain consistent.

\subsection{Proposed Improvements}\label{sec:results-improvements}

\begin{table}
    \caption{Results for the proposed Hybrid/Parallel OCSP variants, compared to the base configurations, for the distributed Earth/Moon/Mars case.}
    \label{tab:improvements-establishment}
    \autobox{
    {\sisetup{round-mode=figures,round-precision=4}
    \begin{tabular}{ccS[table-format=3.2]S[table-format=3.1]S[table-format=2.3]S[table-format=4.1]}
        \toprule
        \multicolumn{2}{c}{Configuration} & \multicolumn{4}{c}{Metrics} \\
        \cmidrule(lr){1-2} \cmidrule(lr){3-6}
        PKI & Variant & {Delivered (\%)} & {Latency} & {Overhead} & {Saturation} \\
        \midrule
        CRL & -- & 99.82 & 318.42 & 24.26 & 991.25 \\
        OCSP & -- & 100.0 & 299.01 & 4.85 & 986.75 \\
        OCSP & Stapling & 100.00 & \best{298.36} & 7.12 & \best{737.25} \\
        OCSP & Validators & 100.00 & \worst{579.55} & \worst{48.73} & \worst{2995.33} \\
        \midrule
        OCSP & Hybrid & 100.00 & 487.62 & 6.59 & 758.50 \\
        OCSP & Parallel & 100.00 & 298.53 & \best{2.04} & 1035.75 \\
        \bottomrule
    \end{tabular}
    }}
\end{table}

Finally, we evaluate each of the techniques proposed in Section~\ref{sec:experiment-design} to see if they measurably improve performance in interplanetary networks, using \framework{} to ensure consistent comparison to our previous results.
All results are given relative to the Earth/Moon/Mars scenario, as these improvements are tailored to the interplanetary setting.

\subsubsection{OCSP Hybrid and OCSP Parallel}

In theory, the OCSP in-transit validation approach should reduce overhead by removing the need for a costly handshake with the CA.
However, we find that in practice this is vastly overshadowed by the negative impact of increased congestion on the links -- sending all traffic via a small number of nodes results in those links becoming severely congested, reducing throughput and limiting performance.
To mitigate this, we propose a hybrid approach which uses in-transit validation for messages traversing the relay links, and normal OCSP Stapling elsewhere.
This retains the benefits of low overhead for cross-segment messages, providing better performance than OCSP Stapling while minimizing the risk of congestion.

\begin{figure}
    \centering\includegraphics[width=.95\linewidth]{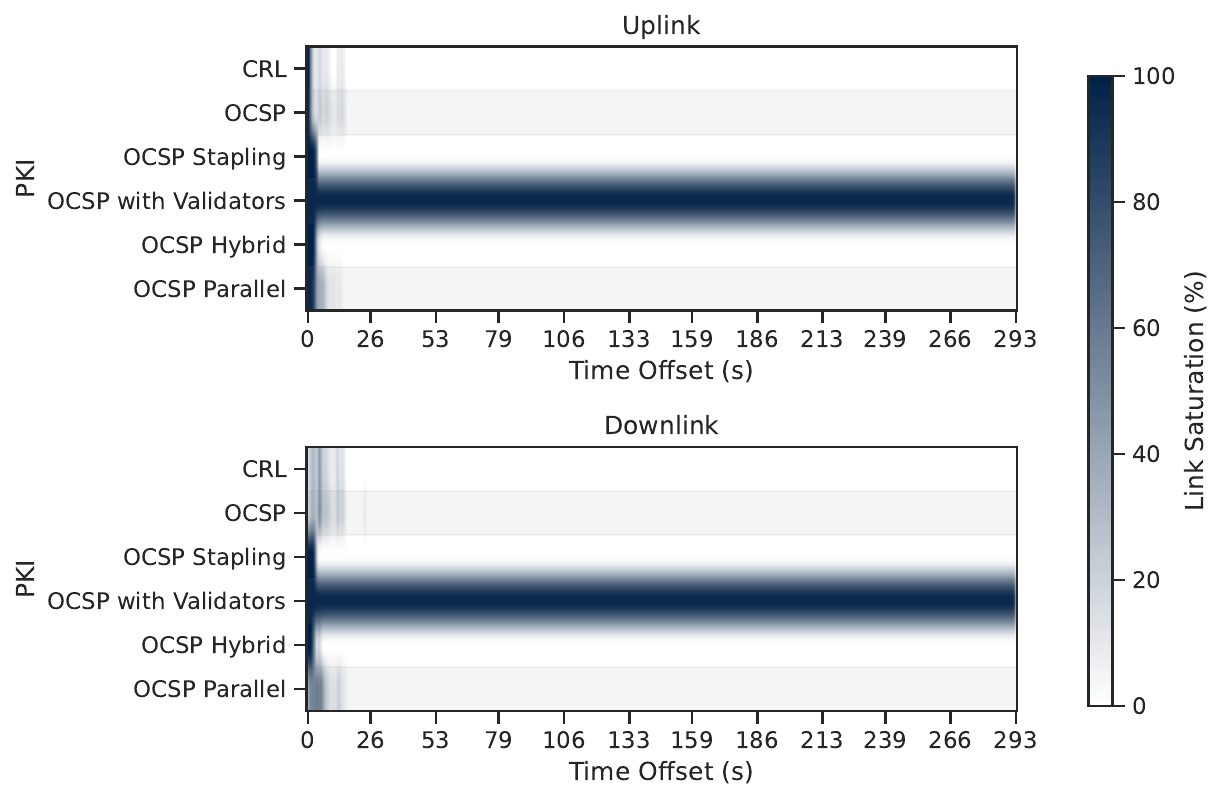}
    \caption{Saturation over time for a link adjacent to one of the CAs in the connection establishment case. Links are saturated for a very long time in the OCSP with Validators case since all messages are sent via the CA -- this is mitigated in the Hybrid and Parallel schemes, which both reduce traffic sent via CAs where possible.}
    \label{fig:ca-saturation}
\end{figure}

We further propose an alternative approach, where the bulk of the message is sent directly to the recipient, with a separate in-transit query sent via the CA. This strategy is aimed at reducing link saturation compared to OCSP with Validators, thus ultimately reducing the risk of congesting the links surrounding the CA.

\begin{table*}
\footnotesize
    \caption{Coverage results for each of our proposed improvements, compared to the base case. Results are taken from the scenario in which attacks are performed on the relay links.}
    \label{tab:improvements-revocation}
    \autobox{
    {\sisetup{round-mode=figures,round-precision=4,detect-weight=true,detect-inline-weight=math}
    \begin{tabular}{cccS[table-format=-4.1]S[table-format=4.0]S[table-format=4.2]S[table-format=4.0]S[table-format=4.1]S[table-format=4.0]}
        \toprule
        & & & \multicolumn{6}{c}{Coverage Time (s)} \\
        \cmidrule(lr){4-9}
        \multicolumn{3}{c}{Configuration} & \multicolumn{2}{c}{No Attack} & \multicolumn{2}{c}{Weakly Attacked ($p=0.1$)} & \multicolumn{2}{c}{Strongly Attacked ($p=0.5$)} \\
        \cmidrule(lr){1-3} \cmidrule(lr){4-5} \cmidrule(lr){6-7} \cmidrule(lr){8-9}
        PKI & Variant & Firewall & {Mean} & {Max} & {Mean} & {Max} & {Mean} & {Max} \\
        \midrule
        CRL & -- & & -104.45 & \worst{1438.80} & 346.80 & \worst{2368.39} & 2670.41 & \worst{6460.43} \\
        OCSP & -- & & -104.45 & \worst{1438.80} & 346.80 & \worst{2368.39} & 2670.41 & \worst{6460.43} \\
        OCSP & Stapling & & \worst{1366.11} & \best{1409.38} & \worst{1671.80} & 2016.18 & \worst{4081.13} & 5722.06 \\
        OCSP & Validators & & \worst{1366.16} & \best{1409.41} & \worst{1671.83} & 2016.20 & \worst{4081.14} & 5722.06 \\
        \midrule
        OCSP & Hybrid & & 1101.38 & \best{1409.28} & 1437.28 & 2082.67 & 3795.01 & 5460.32 \\
        OCSP & Parallel & & 1214.10 & \best{1409.13} & 1593.28 & 2102.74 & 3896.90 & 5042.86 \\
        OCSP & Hybrid & \cmark & -105.50 & 1437.51 & 44.18 & 1886.36 & \best{681.58} & \best{3560.88} \\
        OCSP & Parallel & \cmark & \best{-106.39} & 1434.86 & \best{43.29} & \best{1883.71} & 695.53 & 3602.83 \\
        \bottomrule
    \end{tabular}
    }}
\end{table*}

\begin{figure*}
    \centering\includegraphics[width=\linewidth]{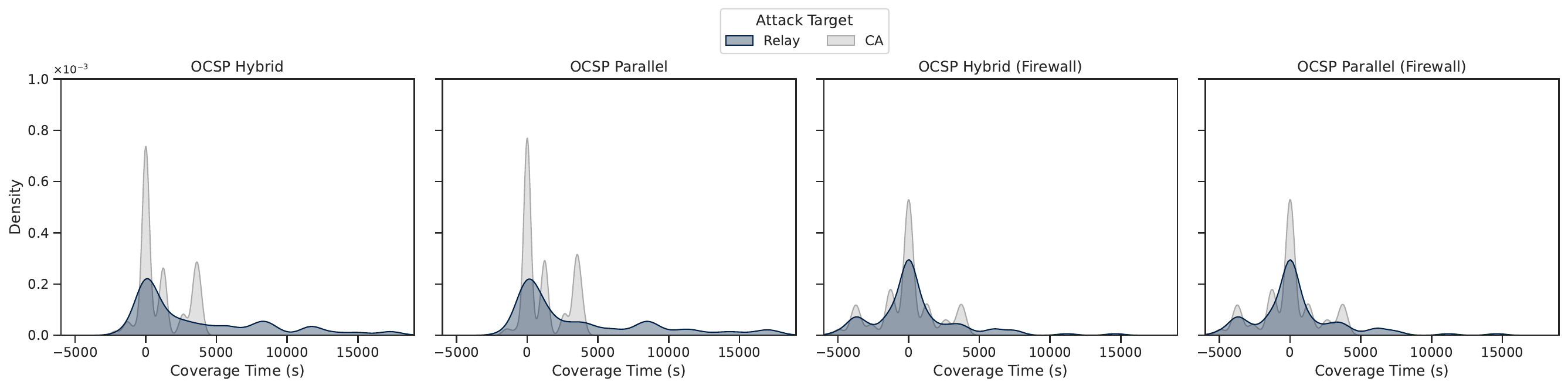}
    \caption{Kernel density estimate for coverage time under each of the proposed improvements, when attackers are jamming relay and CA links with a per-message error rate of $\mathbf{0.5}$. Even under significant jamming, the firewall keeps coverage time low by preventing attacker messages from traversing cross-segment links.}
    \label{fig:attack-saturation}
\end{figure*}

Table~\ref{tab:improvements-establishment} shows that both approaches perform significantly better than OCSP with Validators, reducing congestion to a level comparable to OCSP Stapling, the best-performing base case.
This is further illustrated in Figure~\ref{fig:ca-saturation}, in which congestion dissipates much more quickly on the links surrounding the CA.
Furthermore, both approaches result in comparable latency and even lower overhead compared to the previously tested OCSP configurations, improving performance to the point where establishment overhead is a negligible component of overall latency in the vast majority of cases.

\subsubsection{Relay Firewall}

Finally, we propose a new technique in which the relay nodes are used as a firewall to filter out messages which use revoked certificates.
Under this configuration, the relay filters all messages passing through which are making use of revoked keys, stopping them from reaching their destination or marking them as invalid (depending on implementation).
This has the potential to reduce the race between the revocation and the attacker's messages when messages are waiting to be forwarded at a relay, allowing all of these messages to be dropped by the firewall instead.
It also allows relays to drop messages from an attacker even if a valid certificate status has already been stapled to the message, reducing the reach of the attacker across long-distance relays.

The results can be seen in Table~\ref{tab:improvements-revocation}, and are visualized in Figure~\ref{fig:attack-saturation}.
We can see that the inclusion of a firewall has a huge impact on the coverage time of revocations, when applied to either the Hybrid or Parallel scheme.
This is due to improvements in the case where the attack and revocation originate from different network segments: the attacker's segment is largely unaffected, but the other segments can be protected substantially earlier by filtering messages.
This brings us closer to the expected best-case performance of revocations under distributed PKI, bounding the attacker's reach to the smallest segment possible and covering the other segments in the network as early as possible.
This technique also reduces load on the interplanetary relays, preventing denial of service by attackers using compromised keys to exhaust the bandwidth of relay links.

Since both Hybrid and Parallel OCSP produce very similar results in the revocation scenario, but the Parallel approach significantly reduces traffic and congestion in the establishment scenario, we recommend OCSP Parallel with firewalls for optimal performance across the board.

%% file: discussion.tex
\section{Discussion}\label{sec:discussion}

Our results have confirmed our hypothesis and demonstrated the feasibility of terrestrial PKI protocols in interplanetary contexts. 
We have also proposed novel adaptations to provide better performance in interplanetary settings and demonstrated their effectiveness.

By providing a set of standardized experiments, \framework{} and our implementations enable operators and regulators to test these in a standardized manner.
When new network topologies are proposed or planned, by space agencies or private organizations, they can be tested against the same experiments and protocols to ensure there are no issues with performance arising from the connectivity of nodes.

Another factor that must be considered is the distribution of configuration changes, particularly if operators are changing settings on the fly to account for current usage or ongoing attacks.
This shares many of the challenges present in certificate revocation: the goal is once again to spread configuration changes across the network as quickly as possible.
We can learn from our results in this area, suggesting a decentralized approach to configuration changes -- updates are first pushed to trusted authorities within each network segment, from which they can then be pushed to the remaining nodes.
This reduces the risk that changes are lost, and enables acknowledgment or guaranteed delivery at the segment level, where messages can be quickly retransmitted, rather than relying on broadcasts across long-distance links.

Related works hypothesize that there is no ``one size fits all'' approach to PKI in satellite networks~\cite{viswanathanArchitecture2016,templinDTN2016}.
Our results confirm this: distributed OCSP with firewalls on each of the relays can provide high performance with fast revocation coverage, but high security applications may prefer centralization, even at the cost of increased latency.
Our results enable satellite operators to make informed decisions about PKI implementation based on their specific requirements: \framework{} can be used to test the specifics of a PKI system against a network topology before launch and deployment, ensuring the desired balance between security and performance can be achieved.
Additionally, newly proposed key management systems can be tested against a wide range of network topologies, establishing performance and security characteristics.

Our results have shown that the newly proposed ``Parallel OCSP'' configuration can display significant performance improvements in satellite settings, presenting a ``best of both worlds'' in which there is no need for an additional handshake on connection establishment, while also reducing congestion on links surrounding CAs.
However, it does raise some novel security challenges which must be addressed prior to deployment.
In particular, operators must take care when deciding how nodes deal with messages that have been signed before the certificate validity information arrives -- if attackers are able to spoof these messages in large quantities, it could cause denial of service through exceeding the available buffer space, as nodes wait for messages which will never arrive.
This may be mitigated by limiting the scenarios in which Parallel OCSP is permitted, or through further alterations to the protocol to enable fallback to base OCSP after a time.
It is important that operators consider these questions when deciding how to deploy PKI, and our simulation framework enables them to test options with a fast turnaround time, so decisions can be made with a proper understanding of their impact.

When it comes to security, satellite operators are likely to have overlapping and occasionally incompatible needs, adding complexity when systems are networked together.
We therefore propose a mixed-protocol system, in which key management systems are designed to work alongside one another.
Both OCSP and CRLs are already highly intercompatible, as they are based upon existing internet technologies, and we have shown in our ``OCSP Hybrid'' protocol that mixed approaches are also possible in an interplanetary system.
This would present the opportunity for organizations to choose PKI requirements on a per-device basis, permitting a high degree of control without sacrificing compatibility.

We must also consider the complexity of implementation of these technologies.
Some works evaluated in Section~\ref{sec:related-work} introduce new protocols; although this enables more significant changes, these new protocols have not been tested as extensively.
By using protocols that have already been standardized and are in wide use across the terrestrial internet, we minimize the risk of security threats in the PKI system itself.
We also benefit from the fact that standards already exist, so standardization bodies such as CCSDS do not need to build new standards from scratch, instead building off what already exists.
This could involve further additions to the ``BPSec'' proposed standard~\cite{birraneBundle2022}, adding new block definitions and behaviors to enable the new functionality.
Moreover, this offers the opportunity to accelerate the timeline for standardization, which is critical when the space industry is moving so quickly -- well-defined standards for interplanetary communication will improve compatibility between systems, making the goal of an interplanetary internet easier to realize.

\subsection{Future Work}\label{sec:future-work}

Alongside standardization efforts, some open questions remain which could be addressed in future work.
This paper focuses on reducing latency overhead and link congestion through protocol optimizations, but leaves the optimization of certificate and message size as an open problem.
This could be addressed by using a different certificate with a smaller size, or by modifying the protocol to involve fewer requests to the CA.

Future research could also test ``optimistic'' protocols -- that is, protocols which initially assume the certificate to be valid, rolling back the state if this is later shown not to be the case -- or protocols like QUIC, which enable communication with zero round-trip-time setup~\cite{iyengarQUIC2021}.
However, this raises additional security concerns: firstly, the system must be designed in such a way that state can be efficiently rolled back without losing other data or leaving artifacts behind.
Many satellites have tight resource constraints, and it is vital that adversaries are prevented from exploiting these resources without proper authorization by opening temporary communication channels with invalid credentials.
If these protocols are to be used, there must still be a mechanism for establishing contracts such that resources are not wasted on unauthorized users.
Similarly, there is scope for evaluation of the scope and limitations for retroactive revocations, considering which actions can be safely rolled back without affecting general operation.

%% file: conclusion.tex
\section{Conclusion}\label{sec:conclusion}

In this paper we have introduced \framework{}: a framework for testing key management systems in satellite networks.
This enables protocol designers to ensure their system is efficient and secure, and allows operators to optimize configurations for their network and specific needs.
We have used \framework{} alongside network simulations to test a number of terrestrial PKI techniques applied to interplanetary networks.
In contrast to current consensus, we have shown that terrestrial PKI can be used effectively in current and future satellite networks, ensuring efficient, low-latency connection establishment by distributing certificate authorities across the network.

We have also proposed and tested a number of new techniques building upon these, using \framework{} to demonstrate their improvement in a concrete, repeatable manner.
In particular, our OCSP Parallel technique provides the substantial latency and overhead benefits of sending messages via a CA, while also minimizing link congestion and its impact.
This is clearly demonstrated in our simulation results, in which we show that the addition of a firewall at relay links can further reduce the reach of an attacker with a compromised key by propagating revocation data more quickly, without any impact on standard operation.
Each of these modifications can be implemented in a backwards-compatible manner to enable interoperability with terrestrial networks.

By understanding the impact of PKI configuration on the overall performance of the system, we can ensure future interplanetary networks achieve appropriate security without sacrificing performance.

%% file: appendices/experiment-configuration.tex
\section{Experiment Configuration}\label{app:experiment-configuration}

\begin{figure}[h!]
    \centering\includegraphics[width=\columnwidth]{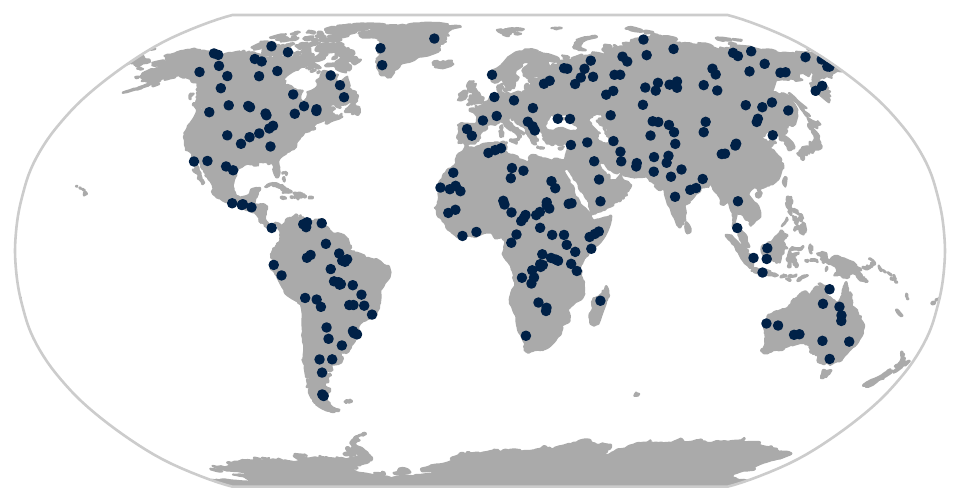}
    \caption{Locations of the ground stations used in the simulations. Positions are randomly generated, but are kept consistent between simulations for repeatability.}
    \label{fig:ground-stations}
\end{figure}

In the experimental framework given in Section~\ref{sec:framework-design}, the terrestrial portion of the network contains 256 ground stations.
These are randomly placed on Earth's land surface, with the positions remaining fixed between simulations.
The positions of the ground stations are visualized in Figure~\ref{fig:ground-stations}; exact coordinates will be included in the experiment configuration code released on publication.

%% file: appendices/protocols.tex
\section{Protocols}\label{app:protocols}

In this appendix we give sequence diagrams for each of the PKI techniques tested in this paper.
These diagrams are shown for the distributed case, in which there are multiple CAs and authority is delegated.
Note that the diagrams are also for nominal operation only -- they do not illustrate behavior in the case of invalid signatures, caching, or revocations.
OCSP Hybrid is not shown here, since behavior is identical to OCSP Stapling for communication within a segment, and identical to OCSP with Validators for communication between segments.

Full protocol specifications are in the simulation source code.

\subsection*{CRL}

\begin{center}
    \resizebox{\linewidth}{!}{
    \begin{tikzpicture}[node distance=3.4cm,auto,>=stealth']
        \node[] (a) {A};
        \node[right = of a] (b) {B};
        \node[right = of b] (ca) {CA\textsubscript{A}};
        \node[below of=a, node distance=4cm] (a_ground) {};
        \node[below of=b, node distance=4cm] (b_ground) {};
        \node[below of=ca, node distance=4cm] (ca_ground) {};
        \draw (a) -- (a_ground);
        \draw (b) -- (b_ground);
        \draw (ca) -- (ca_ground);
        \draw[->] ($(a)!0.285!(a_ground)$) -- node[above,scale=1,midway]{SignedMessage} ($(b)!0.285!(b_ground)$);
        \draw[->] ($(b)!0.571!(b_ground)$) -- node[above,scale=1,midway]{CRLRequestMessage} ($(ca)!0.571!(ca_ground)$);
        \draw[->] ($(ca)!0.857!(ca_ground)$) -- node[above,scale=1,midway]{CRLDirectMessage} ($(b)!0.857!(b_ground)$);
    \end{tikzpicture}
    }
\end{center}

\subsection*{OCSP}

\begin{center}
    \centering
    \resizebox{\linewidth}{!}{
    \begin{tikzpicture}[node distance=3.4cm,auto,>=stealth']
        \node[] (a) {A};
        \node[right = of a] (b) {B};
        \node[right = of b] (ca) {CA\textsubscript{B}};
        \node[below of=a, node distance=4cm] (a_ground) {};
        \node[below of=b, node distance=4cm] (b_ground) {};
        \node[below of=ca, node distance=4cm] (ca_ground) {};
        \draw (a) -- (a_ground);
        \draw (b) -- (b_ground);
        \draw (ca) -- (ca_ground);
        \draw[->] ($(a)!0.285!(a_ground)$) -- node[above,scale=1,midway]{SignedMessage} ($(b)!0.285!(b_ground)$);
        \draw[->] ($(b)!0.571!(b_ground)$) -- node[above,scale=1,midway]{CertificateQueryMessage} ($(ca)!0.571!(ca_ground)$);
        \draw[->] ($(ca)!0.857!(ca_ground)$) -- node[above,scale=1,midway]{CertificateStatusMessage} ($(b)!0.857!(b_ground)$);
    \end{tikzpicture}
    }
\end{center}

\subsection*{OCSP Stapling}

\begin{center}
    \centering
    \resizebox{\linewidth}{!}{
    \begin{tikzpicture}[node distance=3.4cm,auto,>=stealth']
        \node[] (a) {A};
        \node[right = of a] (b) {B};
        \node[left = of a] (ca) {CA\textsubscript{A}};
        \node[below of=a, node distance=4cm] (a_ground) {};
        \node[below of=b, node distance=4cm] (b_ground) {};
        \node[below of=ca, node distance=4cm] (ca_ground) {};
        \draw (a) -- (a_ground);
        \draw (b) -- (b_ground);
        \draw (ca) -- (ca_ground);
        \draw[->] ($(a)!0.285!(a_ground)$) -- node[above,scale=1,midway]{CertificateQueryMessage} ($(ca)!0.285!(ca_ground)$);
        \draw[->] ($(ca)!0.571!(ca_ground)$) -- node[above,scale=1,midway]{CertificateStatusMessage} ($(a)!0.571!(a_ground)$);
        \draw[->] ($(a)!0.857!(a_ground)$) -- node[above,scale=1,midway]{\parbox{5cm}{\centering StapledMessage\\( SignedMessage,\\CertificateStatusMessage )}} ($(b)!0.857!(b_ground)$);
    \end{tikzpicture}
    }
\end{center}

\subsection*{OCSP with Validators}

\begin{center}
    \resizebox{\linewidth}{!}{
    \begin{tikzpicture}[node distance=3.4cm,auto,>=stealth']
        \node[] (a) {A};
        \node[right = of a] (ca) {CA\textsubscript{Relay}};
        \node[right = of ca] (b) {B};
        \node[below of=a, node distance=2.857cm] (a_ground) {};
        \node[below of=b, node distance=2.857cm] (b_ground) {};
        \node[below of=ca, node distance=2.857cm] (ca_ground) {};
        \draw (a) -- (a_ground);
        \draw (b) -- (b_ground);
        \draw (ca) -- (ca_ground);
        \draw[->] ($(a)!0.4!(a_ground)$) -- node[above,scale=1,midway]{\parbox{5cm}{\centering ValidatorQueryMessage\\( SignedMessage )}} ($(ca)!0.4!(ca_ground)$);
        \draw[->] ($(ca)!0.8!(ca_ground)$) -- node[above,scale=1,midway]{\parbox{5cm}{\centering StapledMessage\\( SignedMessage,\\CertificateStatusMessage )}} ($(b)!0.8!(b_ground)$);
    \end{tikzpicture}
    }
\end{center}

\subsection*{OCSP Hybrid}

\paragraph{Inter-segment messages:} See OCSP with Validators.

\paragraph{Intra-segment messages:} See OCSP Stapling.

\subsection*{OCSP Parallel}

\paragraph{Inter-segment messages:} See OCSP with Validators.

\paragraph{Intra-segment messages:} A signed dummy message with minimum size is sent via the CA in a manner consistent with ``OCSP with Validators'', and at the same time the signed real message is sent directly to the recipient.
Once the stapled message arrives from the CA, the recipient can verify the certificate's validity.

\begin{center}
    \resizebox{\linewidth}{!}{
    \begin{tikzpicture}[node distance=3.4cm,auto,>=stealth']
        \node[] (a) {A};
        \node[right = of a] (ca) {CA\textsubscript{A}};
        \node[right = of ca] (b) {B};
        \node[below of=a, node distance=3.2cm] (a_ground) {};
        \node[below of=ca, node distance=3.2cm] (ca_ground) {};
        \node[below of=b, node distance=3.2cm] (b_ground) {};
        
        \draw (a) -- (a_ground);
        \draw (b) -- (b_ground);
        \draw (ca) -- (ca_ground);
        
        \draw[->] ($(a)!0.25!(a_ground)$) -- node[above,scale=1,pos=0.75]{\parbox{5cm}{\centering \textcolor{cbred}{SignedMessage} }} ($(b)!0.25!(b_ground)$);
        \draw[->] ($(a)!0.3!(a_ground)$) -- node[below,scale=1,midway]{\parbox{5cm}{\centering ValidatorQueryMessage\\( \textcolor{cbsky}{SignedMessage} )}} ($(ca)!0.3!(ca_ground)$);
        \draw[->] ($(ca)!0.85!(ca_ground)$) -- node[above,scale=1,midway]{\parbox{5cm}{\centering StapledMessage\\( \textcolor{cbsky}{SignedMessage},\\CertificateStatusMessage )}} ($(b)!0.85!(b_ground)$);

      \draw[fill=cbsky]  (1.0,-3.5) rectangle ++(0.4,0.25);
      \node[anchor=west] at (1.45,-3.4) {\footnotesize Dummy Message};
      \draw[fill=cbred]   (5.0,-3.5) rectangle ++(0.4,0.25);
      \node[anchor=west] at (5.45,-3.4) {\footnotesize Real Message};

    \end{tikzpicture}
    }
\end{center}

%% file: appendices/data-processing.tex
\section{Data Processing}\label{app:data-processing}

In Section~\ref{sec:log-processing} we describe how the \sysname{} logs are extracted and processed to produce the results later in the paper.
Table~\ref{tab:revocation-formulas} describes how the coverage time for a revocation is computed for each PKI approach, taking into account the order of messages in each approach to figure out at what point the victim node is considered protected by a revocation.

\begin{table*}
    \caption{Formulas to compute revocation time for each PKI configuration.}
    \label{tab:revocation-formulas}
    \autobox{
    \begin{tabular}{llll}
    \toprule
    PKI & Variant & Case & Coverage Time Formula \\
    \midrule
    CRL & -- & -- & $D(\mathrm{CA}(N)) - d(A, N) - d(N, \mathrm{CA}(N)) - T_{\mathrm{attack}}$ \\
    OCSP & -- & -- & $D(\mathrm{CA}(N)) - d(A, N) - d(N, \mathrm{CA}(N)) - T_{\mathrm{attack}}$ \\
    OCSP & Stapling & -- & $D(\mathrm{CA}(A)) - d(A, \mathrm{CA}(A)) - T_{\mathrm{attack}}$ \\
    OCSP & Validator & -- & $D(\mathrm{CA}(A)) - d(A, \mathrm{CA}(A)) - T_{\mathrm{attack}}$ \\
    OCSP & Hybrid & Intra-Segment & $D(\mathrm{CA}(A)) - d(A, \mathrm{CA}(A)) - T_{\mathrm{attack}}$ \\
    OCSP & Hybrid & Cross-Segment & $D(R(A,N)) - d(A, R(A,N)) - T_{\mathrm{attack}}$ \\
    OCSP & Hybrid-Parallel & Intra-Segment & $D(\mathrm{CA}(A)) - d(A, \mathrm{CA}(A)) - T_{\mathrm{attack}}$ \\
    OCSP & Hybrid-Parallel & Cross-Segment & $D(R(A,N)) - d(A, R(A,N)) - T_{\mathrm{attack}}$ \\
    \bottomrule
    \end{tabular}
    }
    \noindent\par\vspace{.3em}
    \autobox{
    \footnotesize
    \begin{tabular}{@{\hskip .03\textwidth}p{.065\textwidth}@{}p{.63\textwidth}@{}}
        $A$: & Attacker node \\
        $V$: & Victim node \\
        $\mathrm{CA}(X)$: & Certificate Authority of node $X$ \\
        $R(X, Y)$: & Relay node for cross-segment revocation from $X$ to $Y$ \\
        $D(X)$: & Delivery time of revocation message to $X$ \\
        $d(X,Y)$: & Routing-based propagation delay from $X$ to $Y$ \\
        $T_{\mathrm{attack}}$: & Time of attack \\
    \end{tabular}
    }
\end{table*}

%% file: appendices/latency-results.tex
\section{Latency Results}\label{app:latency-results}

\begin{figure*}
    \centering\includegraphics[width=\linewidth]{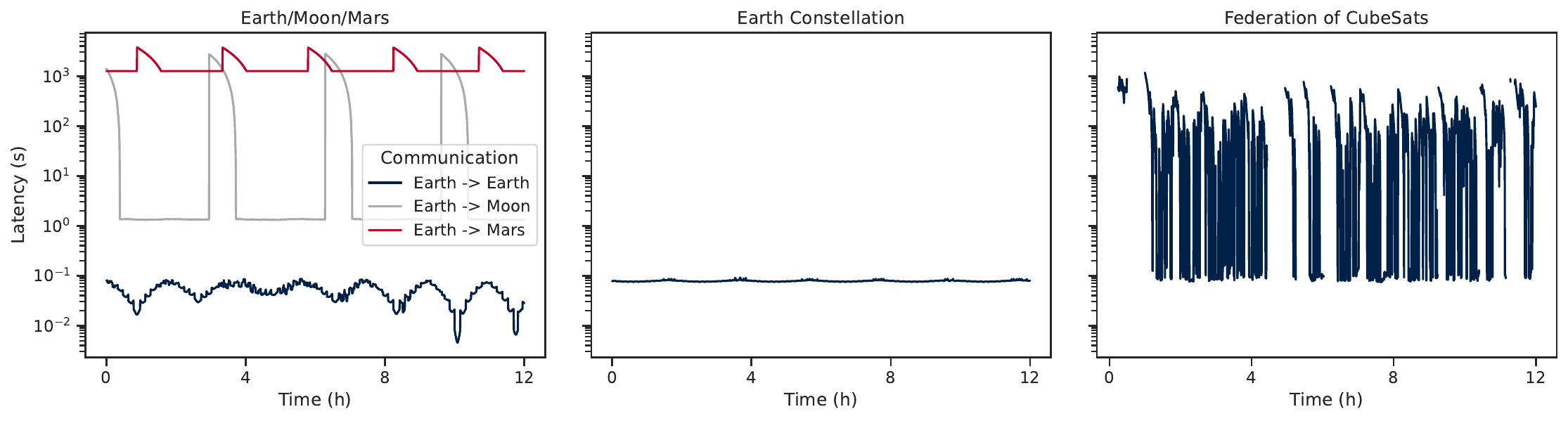}
    \caption{Latency of messages between ground stations on opposite sides of the globe for each of the tested networks.}
    \label{fig:latency-constellations}
\end{figure*}

In this appendix we include additional results on the point-to-point latency of each of the networks tested in this paper.
Figure~\ref{fig:latency-constellations} illustrates how the latency changes over time for each network.
In the Earth/Moon/Mars case we include three elements to show how the latency of communication within the Earth network is much lower than the latency between planetary segments.

%% file: appendices/revocation-coverage.tex
\section{Revocation Coverage}\label{app:revocation-coverage}

\begin{figure*}[t]
    \centering\includegraphics[width=.95\linewidth]{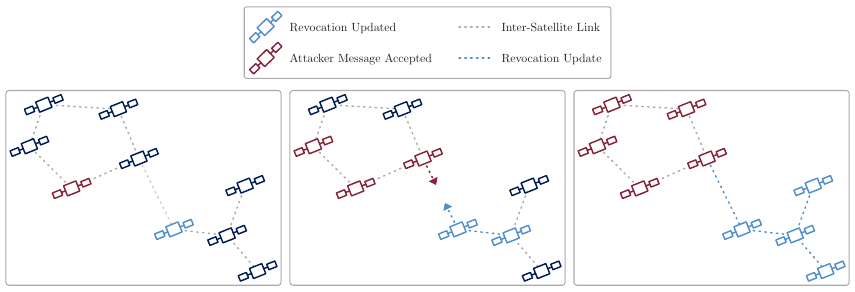}
    \caption{Example of the race between attacker and CA, when an attack and revocation originate in different network segments.}
    \label{fig:revocation-diagram}
\end{figure*}

In this appendix we provide further explanation of the ``coverage time'' results given for the revocation experiments in Section~\ref{sec:results-revocation}.

Figure~\ref{fig:revocation-diagram} illustrates the race between an attacker's message and a revocation, when they originate from different segments.
Due to the vast distances involved, each segment is covered rapidly by its respective message, but cross-segment messages experience a significant delay.
Consequently, the attacker must send their message well in advance of the revocation to ensure it is accepted, as the revocation message will reach the victim CA much more quickly than the attacker's message.
This can result in a negative coverage time, where messages sent before the revocation are rejected in the segment from which the revocation originates, while messages sent after the revocation are still accepted in the attacker's segment.

%% file: appendices/parameter-selection.tex
\section{Parameter Selection}\label{app:parameter-selection}

In Table \ref{tab:parameters} we describe the parameter values we selected for our simulations.
We now explain these choices.
The LTP retransmission limit is defined minimally such that all messages are delivered and is within the NASA recommended range for DTNs~\cite{alma99901077921001842}.
Meanwhile, the LTP retransmission timeout is defined to avoid costly premature retransmissions~\cite{rfc5325-retransmission-function-definition}.
It uses the exact format and values recommended in RFC5325~\cite{rfc5325-retransmission-function-definition}, which motivates the design of LTP.
We lastly set the maximum segment size to \qty{8}{\kilo\byte} because it yields some of the best performance in DTNs while keeping the fragmentation overhead low \cite{11068517}. 

Bandwidth is configured to be a consistent value for each link type, to avoid affecting results.
Many specifications, particularly for future interplanetary constellations, are also not yet publicly available.
We therefore set the per-link bandwidth to \qty{100}{\mega\bit/\second}, a representative value of the wide range of demonstrated link bandwidths in existing and emerging satellite systems~\cite{yangInterSatellite2025, starlink, nasaSOA2024}.
We then set the base message size to \qty{100}{\kilo\byte}.
This is a standard bundle size~\cite{11068517, network3010009} that is also commonly used in related work on space communications \cite{10795038, CAINI202123}.

Finally, we model PKI certificates and certificate revocation lists using the latest versions defined in the X.509 standard~\cite{rfc5280}.
To estimate the sizes of PKI messages, we first trace their protocol-defined structures in the relevant RFCs~\cite{rfc2986,rfc3279,rfc4210,rfc4211,rfc5280,rfc6960}.
We then use example hex dumps in RFC5280~\cite{rfc5280}, together with recommended field sizes, to compute representative message sizes.
To model message signatures, we use RSA-2048, which NIST recommends for use in digital signatures through 2030~\cite{NIST-RSA}.
It produces a \num{256}~byte signature~\cite{rfc8017}, which together with its required algorithm identifier~\cite{rfc5280} results in a signature overhead of \num{275}~bytes.

\begin{table*}
  \caption{Simulation parameters and protocol configuration values used in the experiments.}
  \label{tab:parameters}
  \centering
  \autobox{
  \begin{tabular}{%
    >{\raggedright\arraybackslash}p{2.0cm}  
    >{\raggedright\arraybackslash}p{3.7cm} 
    >{\raggedright\arraybackslash}p{9cm}   
    >{\raggedright\arraybackslash}p{2.6cm}}
    \toprule
    {Type} & {Parameter} & {Description} & {Selected Value} \\
    \midrule
    LTP
      & Retransmission Limit
      & Maximum number of segment retransmissions.
      & \num{15} \\
    LTP
      & Retransmission Timeout
      & Function defining the timeout duration (seconds) for timer-based segments.
      & \makecell[tl]{\texttt{TIMEOUT(RTT)}\\\texttt{= RTT + 2\,$\cdot$\,2}} \\
    LTP
      & Maximum Segment Size
      & Maximum size of any segment.
      & \qty{8}{\kilo\byte} \\
    Transmission
      & Per-Link Bandwidth
      & Unidirectional bandwidth for each link in the network.
      & \qty{100}{\mega\bit/\second} \\
    Transmission
      & Base Message Size
      & Default payload size for messages, without added PKI protocol overhead.
      & \qty{100}{\kilo\byte} \\
    PKI
      & Message Signature
      & Overhead in terms of size from attaching a digital signature to the message.
      & \num{275}~bytes \\
    PKI
      & Certificate Status Message
      & Size of an OCSP response payload containing certificate validity.
      & \num{142}~bytes \\
    PKI
      & CRL Announcement
      & Size of a CRL announcement message payload, where $N$ is the number of revoked certificates in the list.
      & $(98 + 33 \cdot N)$~bytes \\
    \bottomrule
  \end{tabular}
  }
\end{table*}